\documentclass[prb,11pt,onecolumn]{revtex4}
\usepackage{amsmath,amsfonts,amssymb,bm,graphicx,times}
\usepackage[usenames]{color}

\begin{document}

\title{Hybrid-Entanglement in Continuous Variable Systems}

\author{C. Gabriel$^{1,2,*}$,  A. Aiello$^{1,2}$, W. Zhong$^{1,2}$,  T. G. Euser$^{1}$, N.Y. Joly$^{2,1}$, P. Banzer$^{1,2}$, M. F\"ortsch$^{1,2}$, D. Elser$^{1,2}$,  U. L. Andersen$^{1,2,3}$, Ch. Marquardt$^{1,2}$, P.\ St.J.\ Russell$^{1,2}$ and G. Leuchs$^{1,2}$}

\affiliation{
$^{1}$ Max Planck Institute for the Science of Light, Guenther-Scharowsky-Str.\ 1, D-91058 Erlangen, Germany \\
$^{2}$ Institute for Optics, Information and Photonics, University Erlangen-Nuremberg, Staudtstr.\ 7/B2, D-91058 Erlangen, Germany \\
$^{3}$ Department of Physics, Technical University of Denmark, 2800 Kongens Lyngby, Denmark \\
$^{*}$ Christian.Gabriel@mpl.mpg.de
}
\maketitle

\textbf{Entanglement is one of the most fascinating features arising from quantum-mechanics and of great importance for quantum information science. Of particular interest are so-called hybrid-entangled states which have the intriguing property that they contain entanglement between different degrees of freedom (DOFs)\cite{zukowski_test_1991,ma_experimental_2009,neves_hybrid_2009}. However, most of the current continuous variable systems only exploit one DOF and therefore do not involve such highly complex states\cite{ou_realization_1992,bowen_experimental_2002,dong_efficient_2007,boyer_entangled_2008,wagner_entanglingspatial_2008}. We break this barrier and demonstrate that one can exploit squeezed cylindrically polarized optical modes to generate continuous variable states exhibiting entanglement between the spatial and polarization DOF. We show an experimental realization of these novel kind of states  by quantum squeezing an azimuthally polarized mode with the help of a specially tailored photonic crystal fiber.}

Entanglement, a fundamental feature of quantum mechanics, is an essential part of continuous variable quantum information protocols \cite{braunstein_quantum_2005}. An extensive amount of research has focused on generating entanglement in one degree of freedom (DOF), inter alia, in the quadrature \cite{ou_realization_1992}, polarization \cite{bowen_experimental_2002,dong_efficient_2007} or spatial field variables \cite{boyer_entangled_2008,wagner_entanglingspatial_2008}. With these states even the generation of multimode entangled beams is possible, which has the potential to simplify quantum communication systems, especially if multiple modes are contained within a single beam \cite{janousek_optical_2009,lassen_continuous_2009}. 

Manipulating the quantum mechanical properties of more than one DOF has already been demonstrated in the discrete-variable regime, where complex states, such as hybrid- and hyper-entanglement \cite{barreiro_generation_2005,kwiat_hyper-entangled_1997,vallone_hyperentanglement_2009, zukowski_test_1991,ma_experimental_2009}  have been thoroughly investigated. The term `hybrid-entanglement' denotes the peculiar property of quantum states to manifest non-classical correlations  between different DOFs of the system itself \cite{neves_hybrid_2009}.  These different DOFs may belong either to different parties of the same quantum system, e.\ g.\ the matter-part and radiation-part of an atom-light interacting system \cite{waks_protocol_2009,liao_preparation_2006}, or to a single-party system, e.\ g.\ the polarization and time-bin DOFs of a single photon \cite{neves_hybrid_2009}. The latter case is denoted \textit{intra-system} hybrid-entanglement. Hyper-entanglement on the other hand refers to the simultaneous entanglement in all the individual DOFs of a quantum system \cite{kwiat_hyper-entangled_1997}.

In spite of the extensive research in the discrete variable regime, the concept of hybrid- and hyper-entangled states in continuous variable systems is barely developed\cite{dos_santos_continuous-variable_2009}. However, fundamental new concepts for quantum systems in the continuous variable regime, such as hybrid-entanglement, are likely to offer similar advantages for a wide range of applications in quantum information processing to those offered by, for example, discrete-variable hybrid-entanglement\cite{fujiwara_performance_2009,barreiro_beatingchannel_2008}. Here we present the generation of continuous variable multimode hybrid-entanglement by squeezing a cylindrically polarized mode. Such systems represent a new class of yet unexplored continuous variable entangled states.

A classical light field is commonly described by its spatial mode function as well as its polarization vector. In most cases, such a description can be carried out in a single mode picture by constructing a proper mode basis in which only one spatial mode and one polarization mode are occupied. However, as will be shown, such a basis transformation is not possible for cylindrically polarized modes. They display nontrivial structural properties which, when investigated in a quantum-mechanical picture, can yield continuous variable hybrid-entanglement. To illustrate this, consider, for example, the \textit{classical} description of a radially polarized  beam of light which can be written as

\begin{equation}
\mathbf{u}_R (x,y,z)= \frac{1}{\sqrt{2}}\left( \hat{x} \, \psi_{01} + \hat{y} \, \psi_{10} \right) \label{radclass},
\end{equation}

where the functions $\psi_{nm}$, $n,m \in \left\{0,1\right\}$ are the first-order Hermite-Gauss paraxial solutions of the scalar wave equation, and $\hat{x}$ and $\hat{y}$ are unit vectors denoting linear polarization along the $x$ and $y$ axes respectively. If we redefine $\left\{\hat{x},\hat{y}\right\} \equiv \left\{\hat{e}_{1},\hat{e}_{2}\right\}$ and  $\left\{\psi_{10},\psi_{01}\right\}\equiv \left\{v_{1},v_{2}\right\}$, one can rewrite equation (\ref{radclass}) as $\mathbf{u}_{R}(x,y,z)=\sum_{n=1}^{2}\sqrt{\lambda_{n}}\hat{e}_{n}v_{n}$ with $\lambda_{1},\lambda_{2} = 1/2$. Although $\mathbf{u}_{R}(x,y,z)$ represents a perfectly classical object, it has the same tensor-product  form of the quantum state of a two-dimensional bipartite maximally entangled system with Schmidt rank\cite{law_analysis_2004} $K=1/\sum_{n=1}^{2}\lambda_{n}^{2}=2$.
 Therefore $\mathbf{u}_{R}(x,y,z)$ is not separable into the product of a spatially uniform polarization vector $\hat{u}$ and a singular function $f(x,y,z)$: $\mathbf{u}_{R}(x,y,z)\neq \hat{u} \cdot f(x,y,z)$, i.\ e.\  the polarization and spatial DOF are not separable. We will refer to this intruiging classical feature as \textit{structural inseparability}.

A closer inspection of equation (\ref{radclass}) also reveals that, still in analogy with maximally entangled quantum states, it is shape-invariant with respect to simultaneous polarization and spatial basis change. For example,  with $\hat{e}_{\pm}=(\hat{x}\pm i\hat{y})/\sqrt{2}$ and $\phi_{\pm}=(\psi_{10} \pm i\psi_{01})/\sqrt{2}$ denoting the circular and orbital-angular-momentum polarization and spatial bases \cite{law_analysis_2004} respectively, equation (\ref{radclass}) can be rewritten as

\begin{equation}
\mathbf{u}_R (x,y,z) = \frac{1}{\sqrt{2}}\left( \hat{e}_{+} \, \phi_{-} + \hat{e}_{-} \, \phi_{+} \right) \label{LGclass}.
\end{equation}

In order to show that quantum hybrid-entangled states can be generated by squeezing a cylindrically polarized optical mode one needs to change from a classical to a quantum-mechanical description of the state. To do this, let us consider the special case of a bright squeezed azimuthally polarized mode, the state that was used in the experiment. However, it should be noted that the unique features contained in this state can also be observed for other cylindrically polarized optical modes and are, furthermore, not limited to bright states but are also present in squeezed vacuum states. The annihilation operator of the azimuthally polarized mode can be written as
\begin{equation}
\hat{a}_A = \frac{1}{\sqrt{2}}\left( -\hat{a}_{x01} + \hat{a}_{y10}  \right). \label{ReAOps1}
\end{equation}

An azimuthally polarized mode (as well as a radially polarized mode) has a Schmidt-rank of 2, i.e.\ that no matter which basis one chooses, be it the Hermite-Gaussian, Laguerre-Gaussian or any other basis, one always needs two modes to describe this state. Here we choose, in analogy with equation (\ref{radclass}), the Hermite-Gaussian basis to investigate the state.

The  annihilation operator in equation (\ref{ReAOps1}) is associated with the coherent azimuthally polarized eigenstate $\lvert \alpha \rangle_A = \hat{D}_A(\alpha) \lvert 0 \rangle$ where the displacement operator $\hat{D}_A(\alpha)=\exp(\alpha \hat{a}_A^\dagger - \alpha^* \hat{a}_A)$,  $\alpha$ being the classical complex amplitude of the field. To squeeze this state an appropriate squeezing operator can be defined $\hat{S}_A(\zeta)=\exp[(\zeta^* \hat{a}_A^2 - \zeta \hat{a}_A^\dagger\/^2)/2]$, $\zeta$ being  a parameter quantifying the amount of squeezing. The bright squeezed azimuthally polarized state is then given by:
  
\begin{eqnarray}\label{state11}
\lvert \alpha ,\zeta \rangle &=&\hat{D}_A(\alpha) \hat{S}_A(\zeta) \lvert 0 \rangle \nonumber \\
&=& \hat{U}_{x01}(-\alpha/\sqrt{2},\zeta/2 ) \hat{U}_{y10}(\alpha/\sqrt{2}, \zeta/2) \hat{S}_{x01,y10}(-\zeta/2)\lvert 0 \rangle.
\end{eqnarray}
We define $\hat{U}_{i}\left(\alpha/\sqrt{2},\zeta/2 \right) = \hat{D}_i( \alpha/\sqrt{2}) \hat{S}_i(\zeta/2)$ for describing the single modes and additionally the two-mode squeezing operator 

\begin{equation}\label{twomodesq}
\hat{S}_{x01,y10}(-\zeta/2) = \exp[( -\zeta^* \hat{a}_{x01} \hat{a}_{y10} + \zeta \hat{a}_{x01}^\dagger  \hat{a}_{y10}^\dagger )/2].
\end{equation}

\begin{figure}
\centering
\includegraphics[width=0.8\textwidth]{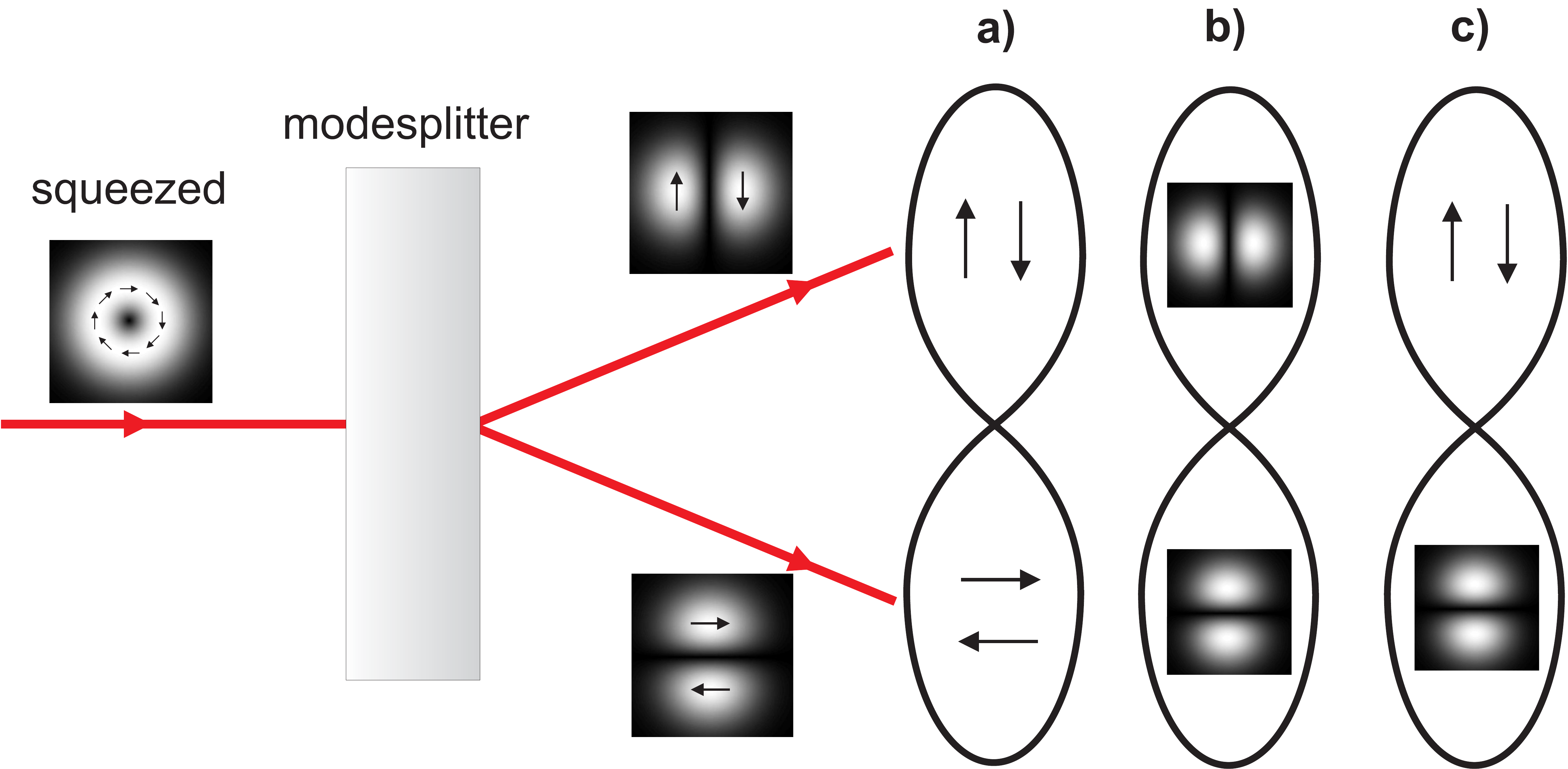}
\caption{\label{theory} \textbf{Illustrating hybrid-entanglement.} Three types of entanglement are contained in a squeezed azimuthally polarized mode which can be observed by utilizing a modesplitter: \textbf{a,} polarization entanglement, \textbf{b,} spatial entanglement and \textbf{c,} hybrid-entanglement. }
\end{figure}

Equation (\ref{state11})  clearly shows that $\lvert \alpha ,\zeta \rangle $ is a two-mode entangled state, where the two-mode squeezing operator (equation (\ref{twomodesq})) fully determines the degree of entanglement of the state. This intra-beam entanglement can be utilized by splitting the cylindrically polarized mode into its two basis modes  with the help of an adequate modesplitter. By making use of the well-known polarization and spatial Stokes parameters\cite{korolkova_polarization_2002,lassen_continuous_2009,hsu_spatial-state_2009} three different kind of measurement sets can be performed on the two modes $\hat{a}_{x01}$ and $\hat{a}_{y10}$ (Fig.\ \ref{theory}):

\begin{enumerate}
	\item Polarization Stokes measurements on mode $\hat{a}_{x01}$ and $\hat{a}_{y10}$.
	\item Spatial Stokes measurements on mode $\hat{a}_{x01}$ and $\hat{a}_{y10}$.
	\item Spatial Stokes measurements on mode $\hat{a}_{x01}$ and polarization Stokes measurements on mode $\hat{a}_{y10}$ or vice versa.
\end{enumerate}

Due to the two-mode squeezing operator (equation ( \ref{twomodesq}))  entanglement between modes $\hat{a}_{x01}$ and $\hat{a}_{y10}$ is present in  all three cases. Especially intriguing is case 3 which shows entanglement between the spatial and polarization DOF. We refer to this entanglement, where  different DOFs are measured in the two arms of the entangled state, as continuous variable hybrid-entanglement in analogy with the discrete variables.  

\begin{figure}
\centering
\includegraphics[width=0.7\textwidth]{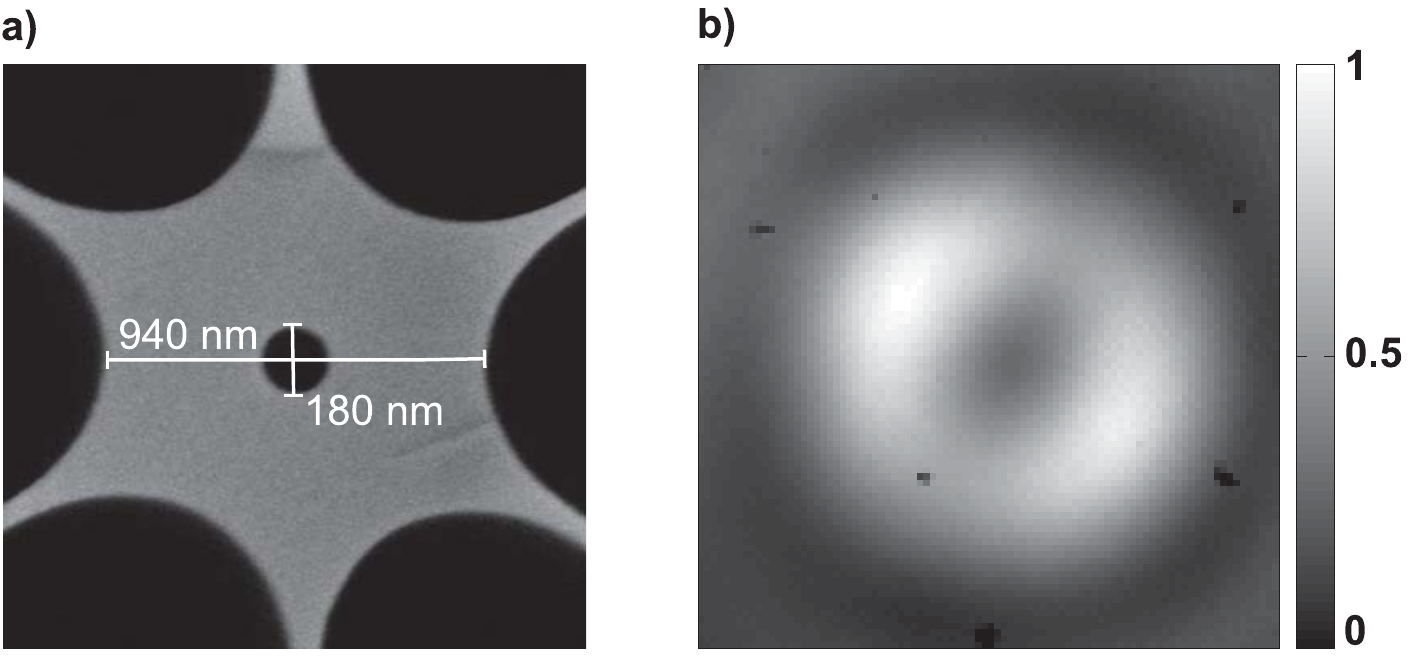}
\caption{\label{Fiber}\textbf{The fiber and the mode.} \textbf{a,} The specially designed photonic crystal fiber which supports the azimuthally polarized mode. \textbf{b,} Normalized mode intensity profile of the azimuthally polarized mode at $\lambda= 810$nm, measured after the fiber.} 
\end{figure}

We experimentally realize such a continuous variable hybrid-entangled state by exploiting the non-linear Kerr effect in a fiber which supports and therefore directly squeezes the azimuthally polarized mode. In the experiment  a mode-locked Ti:sapphire laser, centered at a wavelength of 810 nm and producing 170 fs pulses, acts as light source. As a nonlinear medium a specially designed photonic crystal fiber with a core diameter of 940 nm and in its center a sub-wavelength hollow channel (diameter=180 nm) is chosen (Fig.\ \ref{Fiber}a). This specific structure allows an azimuthally polarized mode to be maintained during propagation, much like the linearly polarized eigenmodes of a polarization-preserving fiber. Similar fibers have already been employed to provide field enhancement in optical fibers\cite{wiederhecker_field_2007}. Furthermore, in this fiber  the azimuthally polarized mode is in the anomalous dispersion regime at a wavelength of 810 nm. Therefore, solitons can form and the fiber is suitable for efficient squeezing generation. A more detailed investigation of the fiber properties is given elsewhere \cite{Euser_Nanobore_2010}. To efficiently excite the desired mode inside the fiber a polarization converter (ARCoptix) is used to generate an azimuthally polarized mode which is then coupled into the fiber. The output mode of the fiber is shown in Fig.\ \ref{Fiber}b. Its asymmetric structure arises from a minor distortion in the cladding structure of the fiber. Therefore the general structure of the core is not perfectly symmetric, leading to an azimuthally polarized mode with slightly higher intensities along one axis.

\begin{figure}
\centering
\includegraphics[width=0.99\textwidth]{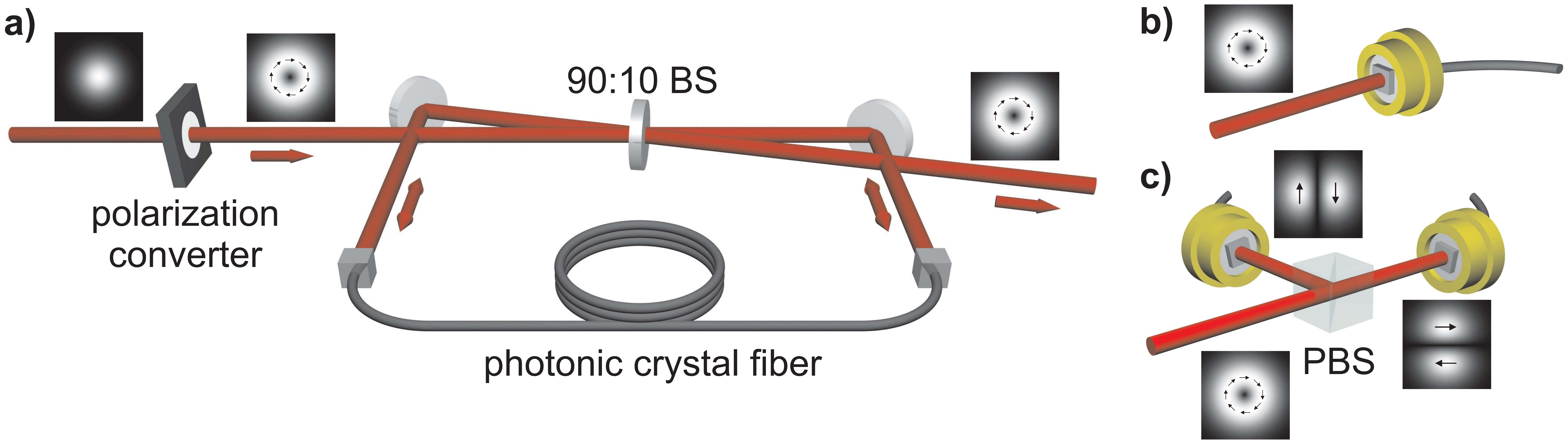}
\caption{\label{Setup}\textbf{Experimental setup.} \textbf{a,} The experimental setup of the Sagnac loop to generate amplitude squeezing in the azimuthally polarized mode; beamsplitter (BS). \textbf{b,} Direct detection is used to measure the amplitude squeezing. \textbf{c,} The detection system to measure amplitude correlations between the horizontally polarized $\textrm{TEM}_{01}$ mode and the vertically polarized $\textrm{TEM}_{10}$ mode; polarizing beamsplitter (PBS).} 
\end{figure}

 In the experimental setup a Sagnac interferometer\cite{schmitt_photon-number_1998}  consisting of a 40 cm long fiber and a beamsplitter with a highly asymmetric splitting ratio of 90:10 is used (Fig.\ \ref{Setup}a). The third-order nonlinear Kerr-effect present in the specially designed photonic crystal fiber  generates quadrature squeezing. However, for certain input energies amplitude squeezing inside the interferometer is generated which can easily be observed with a direct detection scheme (Fig.\ \ref{Setup}b) at the output of the Sagnac-loop. 
   
\begin{figure}
\centering
\includegraphics[width=0.85\textwidth]{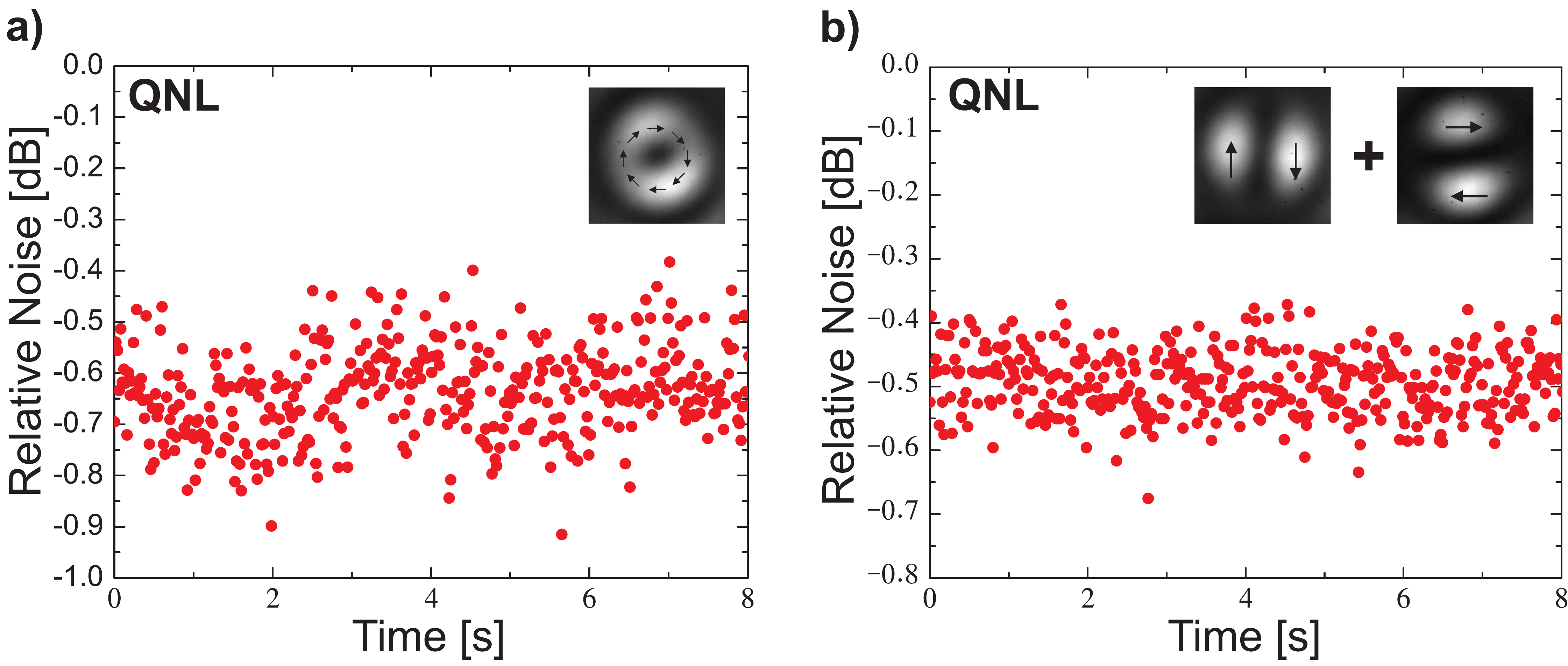}
\caption{\label{SqCor}\textbf{Squeezing and anti-correlations in the azimuthally polarized mode.} \textbf{a,} A quantum noise reduction of 0.6 dB is observed in the azimuthally polarized mode. \textbf{b,} Anti-correlations of 0.5 dB below the QNL of the horizontally polarized $\textrm{TEM}_{01}$ and vertically polarized $\textrm{TEM}_{10}$ have been measured. The insets illustrate the measured spatial modes and the arrows indicate the polarization of the state.} 
\end{figure}

The detection system consists of a detector with sub-shot noise resolution at a radio frequency side band at 10.7 MHz and a high quantum efficiency silicon photodiode. A coherent beam is used to determine the quantum noise limit (QNL). We observe amplitude squeezing of the azimuthally polarized  mode of $0.6\pm0.1$ dB below the QNL (Fig.\ \ref{SqCor}a). The squeezing was observed at pulse energies of about 14 pJ, slightly above the fibers soliton energy.

If the azimuthally polarized  mode is squeezed, equation (\ref{state11}) states that the  horizontally polarized $\textrm{TEM}_{01}$ mode and the vertically polarized $\textrm{TEM}_{10}$ mode are entangled. One indication of this entanglement is the anti-correlation between the modes stated in equation (\ref{ReAOps1}). These can be measured with the detection system illustrated in Fig.\ \ref{Setup}c. It consists of a polarizing beamsplitter (PBS) and two intensity detectors with carefully balanced amplifiers. The PBS acts as a modesplitter, dividing the azimuthally polarized  mode into a horizontally polarized $\textrm{TEM}_{01}$ mode and a vertically polarized $\textrm{TEM}_{10}$ mode which impinge on the two detectors.  By taking the sum signal of the detectors  the anti-correlations are determined, while the QNL is measured with a coherent beam. Anti-correlations of $0.5 \pm 0.1$ dB below the QNL have been observed (Fig. \ref{SqCor}b). This strongly indicates the existence of entanglement between the  horizontally polarized $\textrm{TEM}_{01}$ mode and the vertically polarized $\textrm{TEM}_{10}$.

In conclusion, a novel kind of state, namely a continuous variable hybrid-entangled state, has been presented. This state has the peculiar attribute that its quantum hybrid-entanglement is based on a structural inseparability of the classical state itself. It has been shown that such a highly complex state can be generated in a remarkably convenient manner by squeezing  a cylindrically polarized optical mode, such as an azimuthally polarized mode. 

\section{Methods}

\subsection{Entanglement in a squeezed cylindrically polarized mode}

A squeezed cylindrically polarized mode contains three different types of entanglement, namely polarization, spatial and hybrid-entanglement. These quantum-correlations can be observed by splitting the beam at an appropriate modesplitter and performing suitable polarization or spatial Stokes measurement on arms $a$ and $b$.  One way of examining this entanglement is by applying the continuous variable Peres-Horodecki criterion for separability \cite{duan_inseparability_2000} to the Stokes parameters.  For a symmetric system, i.\ e.\ $\langle \Delta \hat{X}^a \Delta \hat{Y}^a  \rangle= \langle \Delta \hat{X}^b \Delta \hat{Y}^b  \rangle$ with $\hat{X}$ and $\hat{Y}$ being two arbitrary but conjugate quadrature operators and the superscripts $a$ and $b$ labeling two spatially separated subsystems, the criterion is given by:

\begin{equation}
0 \leq V(\hat{S}_{\mu,DOF1}^{a}+\hat{S}_{\mu,DOF2}^{b})+V(\hat{S}_{\nu,DOF1}^{a}-\hat{S}_{\nu,DOF2}^{b}) <4 \left| \alpha \right|, \label{entcritmeth}
\end{equation}

 where $V$ is the standard variance of an operator  $V(\hat{X}) = \langle \hat{X}^2\rangle -\langle \hat{X}\rangle^2$ and $\alpha = \text{cov}(\hat{S}^a_\mu,\hat{S}^a_\nu)= \text{cov}(\hat{S}^b_\mu,\hat{S}^b_\nu)$ is the covariance of two Stokes parameters.  The inequality (\ref{entcritmeth}) can be evaluated for the three combinations of Stokes parameters $( \mu, \nu ) = (1,2)$, $( \mu, \nu ) = (1,3)$, and $( \mu, \nu ) = (2,3)$. By measuring the appropriate combination of polarization or spatial Stokes parameters in arm $a$ or $b$, either polarization $((DOF1,DOF2)=(pol,pol))$, spatial $((DOF1,DOF2)=(spa,spa))$ or hybrid-entanglement $((DOF1,DOF2)=(spa,pol)$ or $(DOF1,DOF2)=(pol,spa))$ can be observed. To perform a polarization Stokes measurement on one arm, the subsystem is combined with an orthogonally polarized coherent beam\cite{korolkova_polarization_2002}. To perform a spatial Stokes measurement, the subsystem is combined with an orthogonal spatial coherent beam\cite{lassen_continuous_2009,hsu_spatial-state_2009}. If, for example, the amplitude of these coherent states is much larger than of the signal states, the Stokes parameters $\hat{S}_{2}$ and $\hat{S}_{3}$ are entangled for all combinations of $(DOF1,DOF2)$:

\begin{equation}
V(\hat{S}_{2,DOF1}^{a}+\hat{S}_{2,DOF2}^{b})+V(\hat{S}_{3,DOF1}^{a}-\hat{S}_{3,DOF2}^{b}) = \;e^{-s}\cosh s <1. 
\end{equation}

Here $V$ is normalized to $4 \left| \alpha \right|$ and the parameter $s$ quantifies the amount of squeezing in the cylindrically polarized mode. If the coherent and entangled states have equal amplitudes, one can show that the Stokes parameter
$\hat{S}_{1}$ and  $\hat{S}_{3}$ are entangled.  This clearly proves that a squeezed azimuthally polarized mode contains polarization, spatial and hybrid-entanglement.

\subsection {Acknowledgments}
The authors thank Annemarie Holleczek and Metin Sabuncu for their advise and support while conducting the experiment. This work was supported by the EU project COMPAS and the Danish Research Council. 

\newpage

\end{document}